\documentclass[11pt,a4paper]{article}
\usepackage[english]{babel}
\usepackage[utf8]{inputenc}
\usepackage{titling}
\usepackage[affil-it]{authblk}
\usepackage{slashed}
\usepackage{epsf}
\usepackage{cite}
\usepackage{graphicx}
\usepackage{subcaption}
\usepackage{verbatim} 
\usepackage{amsmath}
\usepackage{amssymb}
\usepackage{mathrsfs}
\usepackage{mathbbol}
\usepackage{mathtools}
\usepackage{amsfonts}
\usepackage{bbding}
\usepackage{array}
\usepackage[
colorlinks=true
,urlcolor=black
,anchorcolor=black
,citecolor=black
,filecolor=black
,linkcolor=black
,menucolor=black
,linktocpage=true
,pdfproducer=medialab
,pdfa=true
]{hyperref}
\usepackage{color} 
\usepackage[table]{xcolor}

\usepackage[left=2cm,right=2cm,top=2cm,bottom=2cm]{geometry}
\usepackage{tikz}
\usetikzlibrary{shapes,arrows,positioning,automata,backgrounds,calc,er,patterns}
\usepackage[compat=1.1.0]{tikz-feynman}

\vspace {1cm}
\allowdisplaybreaks
\newcommand{\be} {\begin{equation}}
\newcommand{\ee} {\end{equation}}
\newcommand{\bea} {\begin{eqnarray}}
\newcommand{\eea} {\end{eqnarray}}

\newcommand{\cA}{{\mathcal A}}

\newcommand{\mM}{\mathcal{M}}
\newcommand{\SM}{\mathrm{SM}}
\newcommand{\kg}{\kappa_{\gamma\gamma}}
\newcommand{\kz}{\kappa_{Z\gamma}}
\newcommand{\dg}{\epsilon_\gamma}
\newcommand{\dz}{\epsilon_Z}

\newcommand{\ggp}{{\hat g^\prime}}

\newcommand{\published}[1]{%
\gdef\puB{#1}}
\newcommand{\puB}{}

\tikzfeynmanset{
  Wboson/.style = {boson, thick},
  GB/.style     = {scalar, thick},
}

\tikzfeynmanset{
    every blob/.style={
    draw=black, thick,
    pattern=north west lines,
    pattern color=gray!60,
    minimum size=1.2cm,
  }
}
\tikzfeynmanset{warn luatex = false}

\newlength{\diagwd}
\newlength{\diagsep}
\begin{document}

\title{\textbf{Higgs pseudo-observables in $e^+e^-\to h\gamma$}}
\author[1,2]{Fausto Boniotti\thanks{fausto.boniotti@uzh.ch}}
\author[2]{Gino Isidori\thanks{gino.isidori@uzh.ch}}
\author[2]{Christiane Mayer\thanks{christiane.mayer@physik.uzh.ch}}
\author[2]{Emanuelle Pinsard\thanks{emanuelle.pinsard@physik.uzh.ch}}

\affil[1]{\small{PSI Center for Neutron and Muon Sciences, 5232 Villigen PSI, Switzerland}}
\affil[2]{\small{Physik-Institut, Universit\"at Z\"urich, CH-8057 Z\"urich, Switzerland}}

\published{\flushright 
\footnotesize{ZU-TH 38/26}
\vskip2cm }

\maketitle

\begin{abstract}
We analyse the process $e^+e^-\to h\gamma$ in the Standard Model (SM) and in generic extensions with heavy new physics, using the framework of Higgs pseudo-observables. We decompose the one-loop amplitude into separately gauge-invariant photon-pole, $Z$-pole, and non-pole contributions. For heavy new physics, we show that the leading deviations from the SM are necessarily encoded in the $h\gamma\gamma$ and $hZ\gamma$ on-shell effective couplings. 
Genuine $e^+e^-h\gamma$ contact interactions arise only at higher orders in a derivative expansion, both in SMEFT and HEFT. Although Higgs decays constrain the magnitudes of these pseudo-observables, they leave their relative sign undetermined. Measurements of $e^+e^-\to h\gamma$ at different FCC-ee energies can resolve this ambiguity. Deviations incompatible with this description would instead point towards relatively light new degrees of freedom, as illustrated by a simplified $Z'$ model.
\end{abstract}

\section{Introduction}
\label{sec:introduction}

A thorough exploration of the Higgs sector is one of the primary goals of present and future experiments in high-energy physics, and precise measurements of the Higgs-boson couplings are a central part of this programme.
Current measurements at the LHC indicate that these couplings are consistent with the corresponding Standard Model (SM) predictions.
Nevertheless, the present accuracy still leaves room for sizable non-standard effects, particularly in loop-induced couplings such as $hZ\gamma$, which are naturally sensitive to new heavy dynamics. Improving the precision of Higgs-coupling measurements, and identifying observables that provide information complementary to Higgs decay rates, is   a central element of the FCC-ee physics programme~\cite{FCC:2025lpp}. 

Among the processes accessible at FCC-ee, the associated production of a Higgs boson and a photon, $e^+e^-\to h\gamma$, has a number of distinctive features. Neglecting tiny contributions induced by the electron Yukawa coupling, this process occurs in the SM only beyond the tree level. Its cross section is therefore small (of order $0.1~\mathrm{fb}$), but its experimental signature is particularly clean. Moreover, the loop suppression of the SM amplitude naturally implies enhanced sensitivity to new-physics (NP) contributions.

The one-loop SM calculation of $e^+e^-\to h\gamma$ has been known for a long time~\cite{Abbasabadi:1995rc,Djouadi1997}.
The amplitude receives contributions from effective $h\gamma\gamma$ and $hZ\gamma$ vertices, mediated predominantly by $W$-boson and top-quark loops, as well as from non-pole contributions dominated by $W$-box diagrams. The interference among these different terms leads to a non-trivial dependence of the cross section on the centre-of-mass energy, with a broad maximum around $240$--$250$~GeV. The process has also been investigated in several extensions of the SM, including supersymmetric theories, extended scalar sectors, and other explicit models~\cite{Kanemura:2018esc,Hung:2019jue,Rahili:2019ixf,He:2020suf}.
Complementary analyses have studied possible deviations through anomalous Higgs couplings or within the Standard Model Effective Field Theory (SMEFT)~\cite{Cao:2015iua}.

In this work, we propose a complementary and largely model-independent perspective on $e^+e^-\to h\gamma$, particularly suited to interpreting possible measurements at FCC-ee. Our analysis is based on the concept of Higgs pseudo-observables
\cite{Gonzalez-Alonso:2014eva,Greljo:2015sla,
LHCHiggsCrossSectionWorkingGroup:2016ypw}, which provides a useful interface between experimental measurements and theoretical interpretations. The pseudo-observables are defined directly from the pole structure of physical amplitudes and provide a characterization of interactions that is independent of the choice of fields and operator basis used in  effective-theory (EFT) approaches. 
At the same time, they can be computed within any specified effective field theory or explicit NP model. 

As a first step, we reconsider the SM amplitude and organize it into three separately gauge-invariant components: a photon-pole contribution, a $Z$-pole contribution, and a non-pole remainder. Schematically, for each chirality of the initial-state leptons, the scalar part of the amplitude takes the form
\begin{equation}
    \mathcal A^{L,R}(s,t)
    =
    \frac{\mathcal{R}_{\gamma}}{s}
    +
    \frac{\mathcal{R}_Z^{L,R}}{s-m_Z^2}
    +
    \mathcal A_{\mathrm{non\text{-}pole}}^{L,R}(s,t)\,.
    \label{eq:intro_amplitude}
\end{equation}
Here, the residues $\mathcal R_\gamma$ and $\mathcal R_Z^{L,R}$ are directly related to the {\em on-shell} $h\to\gamma\gamma$ and $h\to Z\gamma$ amplitudes, respectively. The non-pole term contains the regular parts of the vertex amplitudes, together with the box contributions. We have repeated the complete SM one-loop calculation, confirming the known results for the cross section and explicitly verifying the gauge independence of each of the three components in Eq.~\eqref{eq:intro_amplitude}. This decomposition provides the natural starting point for discussing new-physics effects.

In generic extensions of the SM in which the new degrees of freedom are heavy compared with the energies probed in the process, the leading deviations from the SM can be separated into two classes. The first corresponds to modifications of the $h\gamma\gamma$ and $hZ\gamma$ pole residues. These are described by the same pseudo-observables that determine the radiative Higgs decays $h\to\gamma\gamma$ and $h\to Z\gamma$. The second class consists of genuine $e^+e^-h\gamma$ contact interactions, which contribute to the non-pole part of the amplitude. This distinction follows from the analytic structure of the amplitude and is therefore general and gauge invariant. In particular, it does not rely on whether electroweak symmetry breaking is described through a linear or a non-linear realization.

An important hierarchy between these two classes of contributions emerges from effective-field-theory power counting. In SMEFT, local interactions modifying the $h\gamma\gamma$ and $hZ\gamma$ pole residues are already generated by operators of canonical dimension six. Genuine $e^+e^-h\gamma$ contact interactions preserving chirality, instead, first arise at dimension eight. An analogous hierarchy is present in the Higgs Effective Field Theory (HEFT): the pole pseudo-observables can be modified by interactions of dimension five, whereas the relevant contact terms first appear at dimension seven. Barring special suppressions of the lower-dimensional contributions, modifications of the two pole residues are therefore expected to provide the dominant effects of heavy new physics in $e^+e^-\to h\gamma$.

The information accessible in this process is not, however, identical to that obtained from radiative Higgs decays. Measurements of $h\to\gamma\gamma$ and $h\to Z\gamma$ constrain the absolute values of the corresponding amplitudes, but do not determine their relative sign. In $e^+e^-\to h\gamma$, the photon- and $Z$-pole amplitudes interfere and their relative weight changes in a well-defined manner dictated by the corresponding propagators. Measurements of the cross section at two or more centre-of-mass energies can thus resolve the sign ambiguity that is intrinsically present in Higgs decay rates. The availability of different FCC-ee energy stages is particularly valuable in this respect: the energy dependence of the process becomes an essential part of the measurement rather than merely a means of increasing its statistical sensitivity.

This observation also leads to a useful diagnostic of possible new-physics signals. Since the magnitudes of the $h\gamma\gamma$ and $hZ\gamma$ amplitudes are already constrained by Higgs decays, and will be determined more precisely in the future, obtaining a large enhancement of $e^+e^-\to h\gamma$ from heavy new physics is generally difficult. One possibility is a change in the relative sign of the two pole residues, leading to a substantial modification of their interference. Such a configuration is possible, but requires a tuned cancellation between the SM and NP contributions. Most importantly, it predicts a characteristic energy dependence that can be tested using measurements at different centre-of-mass energies.

If an observed enhancement cannot be described by values of the pole pseudo-observables compatible with Higgs decay data, the assumptions underlying the heavy-new-physics expansion must be reconsidered. In particular, such a signal could indicate the presence of additional non-local structures associated with new degrees of freedom not far above the experimentally accessible energy range. To illustrate this possibility, we consider a simple scenario involving a new neutral vector boson with sizable effective coupling to the Higgs and photon fields. 
This example is not intended as a complete ultraviolet model, but rather as an explicit demonstration of the rather special
type of new physics that could generate an energy dependence that cannot be reproduced by shifts of the two Higgs pseudo-observables alone.

The paper is organized as follows. In Sec.~\ref{sec:amplitude}, we discuss the SM amplitude and its decomposition into gauge-invariant pole and non-pole terms, and present the corresponding cross section. In Sec.~\ref{sec:heavyNP}, we formulate the heavy-NP contributions in terms of the $h\gamma\gamma$ and $hZ\gamma$ pseudo-observables and analyze the information provided by measurements at different FCC-ee energies. Section~\ref{sec:beyond-kappa} discusses effects that cannot be captured by the two Higgs pseudo-observables, and we summarize our conclusions in the final section.

\section{Anatomy of amplitude and cross-section within the SM}
\label{sec:amplitude}
\subsection{Amplitude decomposition}
Within the SM, the tree-level amplitude for $e^+ (p_+) e^- (p_-) \rightarrow \gamma (k) h (p_h)$ is suppressed by the tiny electron Yukawa coupling. In the massless-electron limit adopted throughout this paper, non-vanishing contributions to this process start at the one-loop level. In this limit chirality is preserved and the amplitude can be decomposed as 
\begin{equation}
  \mM=\mM^L_{\mu\nu}\,
  \bar{v}(p_+) \gamma^\mu P_L u(p_-) 
  \epsilon_\gamma^\nu +
  \mM^R_{\mu\nu}\,
  \bar{v}(p_+) \gamma^\mu P_R u(p_-) 
  \epsilon_\gamma^\nu \,.
\end{equation}
Gauge and Lorentz invariance further imply 
\begin{align}
\mathcal{M}_{\mu\nu}^{L,R} & = \mathcal{A}^{L,R}_q (s,t)\, \mathcal{T}_{\mu\nu}(q,k) +  \mathcal{A}^{L,R}_d (s,t)\, \mathcal{T}_{\mu\nu}(d,k)\,,
\label{eq:chiral:tensor:qd} 
\end{align}
where 
\begin{align}
 \mathcal{T}_{\mu\nu}(p,k)= g_{\mu\nu}(p\cdot k)-k_\mu p_\nu\,,
\end{align}
and we defined $q = p_+ + p_-$ and $d=  p_- - p_+$. The 
 kinematical variables
\begin{equation}
s=(p_++p_-)^2\,,
\quad t=(p_--k)^2,\quad u=(p_+-k)^2,
\end{equation}
satisfy $s+t+u=m_h^2$.

\begin{figure}[t]
\setlength{\diagwd}{0.47\textwidth}
    \begin{subfigure}[b]{\diagwd}
    \centering
    \includegraphics[]{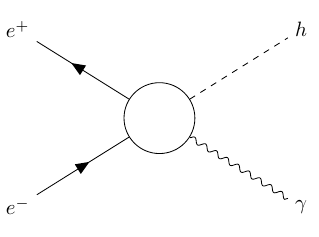}
    \subcaption{}
 \label{diag:eff:4}
  \end{subfigure}\hfill
  \begin{subfigure}[b]{\diagwd}
  \centering
   \includegraphics[]{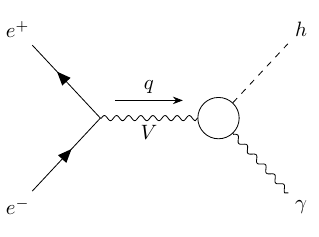}
   \subcaption{}
   \label{diag:eff:s}
    \end{subfigure}
    \bigskip
    \begin{subfigure}[b]{\diagwd}
    \centering
    \includegraphics[]{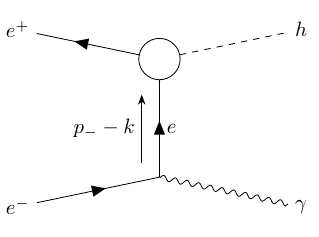}
    \subcaption{}
    \label{diag:eff:t}
    \end{subfigure}
    \hfill
    \begin{subfigure}[b]{\diagwd}
    \centering
    \includegraphics[]{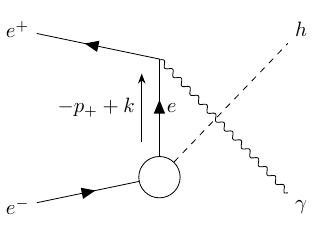}
    \subcaption{}
    \label{diag:eff:u}
    \end{subfigure}
  \caption{Feynman-diagram topologies relevant to the 
  $e^+  e^-\rightarrow \gamma h$ transition in the SM: (a) indicates  the one-particle irreducible (1PI) four-point function; (b) denotes the one-particle reducible diagrams with irreducible 
  three-point functions (effective vertices) with single-particle exchange in the $s$ channel ($V=\gamma,Z$); (c) and (d) indicate the effective vertices associated to singularities in the $t$ and  $u$ channels, respectively.}
  \label{fig:effective:vertices}
\end{figure}

The relevant Feynman-diagram topologies are illustrated in Figure \ref{fig:effective:vertices}. We denote by $\mM_\gamma$ ($\mM_Z$) the $s$-channel diagrams (Fig.~\ref{fig:effective:vertices}b) due to the exchange of a virtual photon ($Z$ boson). These classes include triangle diagrams with fermions and $W$ bosons, as well as quartic couplings of gauge bosons. 
Diagrams with $Z$-$\gamma$ mixing are also included in $\mM_Z$. The remaining terms are box diagrams contributing to the effective 4-point vertex (Fig.~\ref{fig:effective:vertices}a) as well as contributions to the $u$- and $t$-channels 
(Fig.~\ref{fig:effective:vertices}c,~\ref{fig:effective:vertices}d). We combine these into $\mM _{W\text{-box}}$ and $\mM_{Z\text{-box}}$ depending on which bosons are contributing inside the loops, so that the total amplitude can be expressed as
\begin{equation}
    \mM = \mM_\gamma + \mM_Z + \mM _{W\text{-box}} + \mM_{Z\text{-box}}.
    \label{eq:Mtot}
\end{equation}
The sums of the diagrams in each of these classes take the form
\begin{align}
    \mM_\gamma &=  \bar{v}(p_+)\left[ e\, Q_e\; \gamma^\alpha D^\gamma_{\alpha\mu}(s)\;i \Gamma_{h \gamma\gamma}^{\mu\nu}(s) \right]u(p_-)\epsilon_\nu^*(k)\,,  \label{eq:Mgamma} \\
    \mM_Z &=\bar{v}(p_+)\left[  \frac{g}{c_W} \gamma^\alpha\left(P_L \,T^3_e -Q_e\, s_W^2\right)D^Z_{\alpha\mu}(s) \;i\Gamma_{h Z\gamma}^{\mu\nu}(s)\right]u(p_-)\epsilon_\nu^*(k)\;, \label{eq:MZ} \\
     \mM_{W \text{-box}} &= \bar{v}(p_+) \gamma^\mu P_L \mathcal{B}^{ W }_{\mu\nu}(s,t,u)   \mathit{u}(p_-) \epsilon^{* \nu} (k)\;, 
     \label{eq:MBW} \\
    \mM_{Z \text{-box}} &= \bar{v}(p_+) \gamma^\mu \frac{1}{c_W^3}  \left(  \frac{1}{2}(1-2s_W^2)^2P_L+2 s_W^4P_R   \right)\mathcal{B}^{Z}_{\mu\nu}(s,t,u) u(p_-) \epsilon^{* \nu} (k)\,,
    \label{eq:MBZ} 
\end{align}
where $D^V_{\alpha \mu}(s)$ are the propagators of the respective vector bosons in the $s$-channel exchange, and $\Gamma^{\mu \nu}_{V h \gamma} (s)$ are the effective 3-point vertices.  The analytic expressions for the effective vertices and the $\mathcal{B}$ functions in the Feynman gauge can be found in the appendix.
These results were derived first in~\cite{Abbasabadi:1995rc,Djouadi1997}. We  cross-checked these findings starting from the calculation of the amplitude in a generic $R_\xi$ gauge, verifying also the gauge-independence of the total result. 

While the sum of all terms in Eq.~(\ref{eq:Mtot}) is gauge-independent, this is not true for the separate terms in Eqs.~(\ref{eq:Mgamma})--(\ref{eq:MBZ}). On the other hand, the physical poles in the s-channel diagrams, namely 
\begin{align}
    \mM_{\gamma\text{-pole}} &= \frac{e Q_e}{s}\; 
    \Gamma_{ h \gamma\gamma}^{\mu\nu}(s=0)\;
    \bar{v}(p_+) \gamma_\mu  u(p_-)\epsilon_\nu^*(k)\,, \\
    \mM_{Z \text{-pole}} &=
    \frac{g}{c_W} \frac{1}{s-m_Z^2}\; \Gamma_{h Z\gamma}^{\mu\nu}(s=m_Z^2)\;
    \bar{v}(p_+)\left[   \gamma_\mu\left(P_L \,T^3_e -Q_e\, s_W^2\right)\right]u(p_-)\epsilon_\nu^*(k)\;, 
\end{align}
are separately gauge invariant. Correspondingly, also the residual non-pole term
\begin{align}
    \mM_\text{res} &= \mM_{W\text{-box}} + \mM_{Z\text{-box}} + \left( \mM_\gamma - \mM_{\gamma \text{-pole}} \right) + \left( \mM_Z - \mM_{Z \text{-pole}} \right)
\end{align}
is gauge invariant. This latter decomposition into gauge-invariant pole and non-pole terms is dictated by the analytic properties of the amplitude and remains valid also for generic heavy new-physics contributions. 

Since the pole amplitudes depend on the sum of the incoming lepton momenta, they only contribute to the scalar amplitude $\mathcal{A}^{L,R}_q$ as defined in Eq.~\eqref{eq:chiral:tensor:qd}. In general, separating pole and non-pole terms, we can write 
\begin{align}
    \mathcal{A}^{L,R}_{q}(s,t)  =
    \frac{\mathcal{R}_{\gamma}}{s}
    +
    \frac{\mathcal{R}_Z^{L,R}}{s-m_Z^2}
    + 
    \mathcal{A}^{L,R}_{q,\, {\rm res}}(s,t)\,,
    \qquad 
    \mathcal{A}^{L,R}_{d}(s,t)  =
    \mathcal{A}^{L,R}_{d,\, {\rm res}}(s,t)\,,
\end{align}
where the residues $\mathcal R_\gamma$ and $\mathcal R_Z^{L,R}$ are reported in the appendix.

\subsection{Cross section}

\begin{figure}[t]
\centering\includegraphics[width=0.7\textwidth]{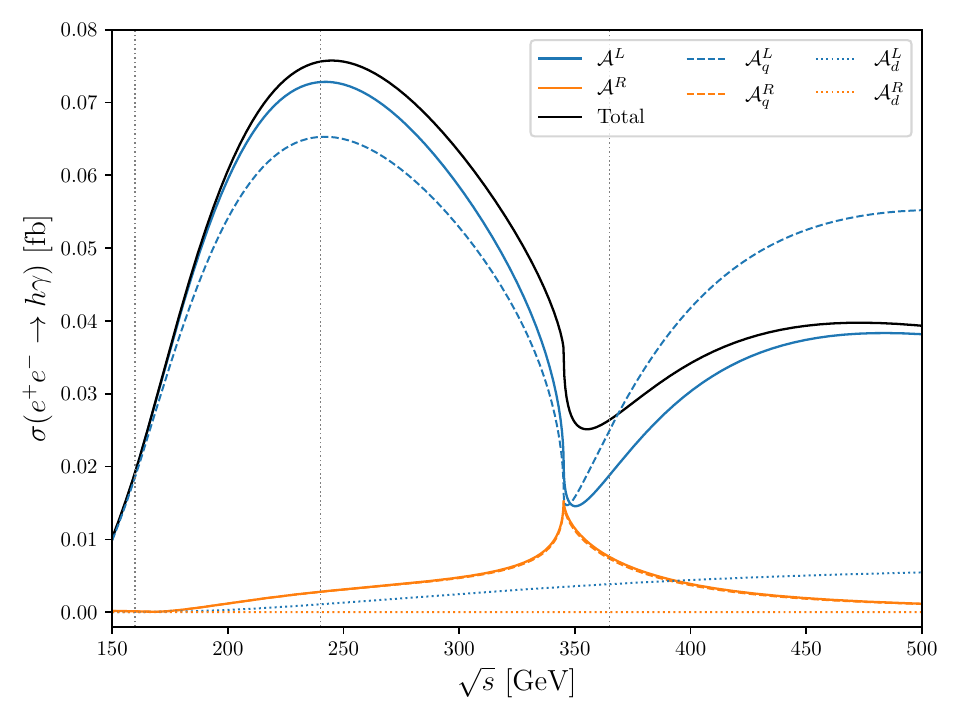}
 \caption{$\sigma(e^+e^- \to h\gamma)$
 as obtained within the SM at the one-loop level.
 The different lines illustrate the contributions of the two chiralities as well as $q$- and $d$-type amplitudes. The numerical results have been obtained using the inputs in Eq.~\eqref{eq:inputs}. Dotted vertical lines indicated reference energies for the FCC-ee runs: $\sqrt{s}=160,\,240$~GeV (stage-I) and $\sqrt{s}=365$~GeV~(stage-II).}
 \label{fig:sm}
\end{figure}

The unpolarized differential cross section reads
\begin{align}
\begin{split}
    \frac{d\sigma}{ d\cos\theta}= \frac{1}{512 \pi}\frac{(s-m_H^2)^3}{s}  &\left\{ \left(1+\cos\theta^2\right) \left[ |\mathcal{A}^L_q (s,t)|^2 + |\mathcal{A}^L_d(s,t)|^2+|\mathcal{A}^R_q(s,t) |^2 + |\mathcal{A}^R_d(s,t)|^2\right] \right. \\
    & \left.  - 4\cos\theta \;\text{Re}  \left[\mathcal{A}^L_q(s,t) \mathcal{A}^L_d (s,t)^*+\mathcal{A}^R_q(s,t) \mathcal{A}^R_d (s,t)^*\right]
    \right\}\,,
\end{split}
\end{align}
where we used $\{t,u\}=-(s-m_h^2)(1\mp \cos\theta)/2$. The second term encodes the interference between the $q$- and $d$-type amplitudes and contributes to the non-trivial angular dependence of the process.
In Figure~\ref{fig:sm} we show the SM expectation as a function of $\sqrt{s}$, separating the two helicities as well as $q$- and $d$-type amplitudes. The plot has been obtained using the following inputs~\cite{ParticleDataGroup:2026mpi}
\begin{align}
 \{m_h,m_W,m_Z,m_t\} = \{125.20,\, 80.37,\, 91.19,\, 172.57\}\ \mathrm{GeV}\,,
 \label{eq:inputs}
\end{align} 
setting further $\alpha^{-1}=137.036$ and  $s_W^2=1-m_W^2/m_Z^2$. 

The two most notable features of Figure~\ref{fig:sm} are
\begin{itemize}
\item the non-trivial $\sqrt{s}$ dependence, with a broad maximum around $240$--$250$~GeV and a destructive-interference region near the $t\bar t$ threshold;
\item the dominance of the left-handed $q$-type component, especially at low $\sqrt{s}$. 
\end{itemize} 

\medskip
At the two reference $\sqrt{s}$ values relevant for FCC-ee stage I, we get
\begin{equation}
 \sigma^{\SM}_{[160]} (h\gamma) =1.96\times 10^{-2}~\mathrm{fb}\,,\qquad
 \sigma^{\SM}_{[240]} (h\gamma) =7.69\times 10^{-2}~\mathrm{fb}\,.
 \label{eq:rates}
\end{equation}
We stress that these are parton-level results that do not include initial-state radiation, higher-order corrections, or kinematical cuts. 
The QCD corrections are small around 240 GeV but become appreciable above the top threshold~\cite{Sang:2017vph}.
Initial-state effects must be included in a precision comparison with data. However, they are expected to cancel to good accuracy in the ratio of the cross section in the presence of generic heavy NP to its SM value, which is the main observable discussed below.

\section{NP effects via Higgs pseudo-observables}
\label{sec:heavyNP}

We now consider extensions of the SM in which the new degrees of freedom are sufficiently heavy that their effects can be organized in a local expansion over inverse powers of a scale $\Lambda_{\rm NP}\gg \sqrt{s},m_h$. In this regime the pole structure of the amplitude is unchanged: NP can modify the residues of the photon and $Z$ poles and can generate additional analytic terms in the kinematic invariants. The relative importance of these effects follows directly from EFT power counting.

A helicity-conserving local contribution to $e^+e^-\to h\gamma$ requires, after electroweak symmetry breaking, an interaction of the schematic form
\begin{equation}
  J_{L,R}^\mu \partial^\nu h F_{\mu\nu}\,,
  \label{eq:contact-structure}
\end{equation}
where $J_{X}^\mu = \bar{e}_X \gamma^\mu e_X$.
In SMEFT, where the physical Higgs boson belongs to an $SU(2)_L$ doublet, the gauge-invariant completion of Eq.~\eqref{eq:contact-structure} first occurs at dimension eight. In HEFT, where $h$ is treated as an electroweak singlet, the interaction in Eq.~\eqref{eq:contact-structure} has dimension seven. By contrast, the $h\gamma\gamma$ and $hZ\gamma$ pole residues can already be modified by dimension-six operators in SMEFT and dimension-five interactions in HEFT. Genuine helicity-conserving contact terms are therefore suppressed by two additional powers of the heavy scale in either expansion.

Dipole interactions of the form $\bar e\sigma^{\mu\nu}e\,hF_{\mu\nu}$ constitute a possible exception to this counting, since they arise at dimension six in SMEFT and dimension five in HEFT. However, they flip the electron chirality hence do not interfere with the SM amplitude in the massless-electron limit. We neglect their quadratic contribution, consistently with our expansion and with the stringent constraints on electron dipole interactions.\footnote{~In SMEFT, electroweak gauge invariance relates these interactions to the electron electromagnetic dipoles, which are tightly constrained by measurements of the electron anomalous magnetic moment and electric dipole moment.}

It follows that, at the leading non-trivial order in the heavy-NP expansion, deviations in $e^+e^-\to h\gamma$ are described by non-standard values of the $h\gamma\gamma$ and $hZ\gamma$ pole residues. The same quantities control the on-shell amplitudes for $h\to\gamma\gamma$ and $h\to Z\gamma$. In SMEFT they can be correlated with other electroweak observables, hence deviations from the SM values are highly constrained. Such correlation does not hold in HEFT, where sizable deviations from the SM are still possible. The relation among the two radiative Higgs decays and $e^+e^-\to h\gamma$, being dictated by the pole structure of physical amplitudes, is instead common to both frameworks.

Restricting for simplicity to CP-even heavy-NP effects, we define two real pseudo-observables at the amplitude level,
\begin{equation}
 \kappa_{V\gamma}=\frac{\mathcal A(h\to V\gamma)}{\mathcal A_{\SM}(h\to V\gamma)}\,,
 \qquad
 \mu_{V\gamma} = \kappa^2_{V\gamma}= 
 \frac{\Gamma(h\to V\gamma)}{\Gamma_{\SM}(h\to V\gamma)}\,.
 \label{eq:mu}
\end{equation}

Decay rates determine $|\kappa_{V\gamma}|$, but are insensitive to its sign.\footnote{~If also CP-violating terms are included, we should distinguish 
CP-even and CP-odd contributions to $\cA(h\to V\gamma)$, where only
CP-even terms interfere with 
the SM amplitude. The formalism we adopt is valid also in this case, provided $\kappa_{V\gamma}$ is understood as the ratio of the CP-even amplitudes and we assume that quadratic effects in the rates 
due to CP-odd NP amplitudes are negligible.}
With two independent radiative channels, decay data alone leave four discrete sign assignments for $\{\kg,\kz\}$.

Defining further $\dg=\kg-1$ and $\dz=\kz-1$, so that the SM limit is recovered for $\dg=\dz=0$, the leading heavy-NP effects in $e^+e^-\to h\gamma$ 
 can be obtained by modifying only the pole part of the amplitude:
\begin{align}
    \mathcal{A}^{L,R}_{q}(s,t)  = \mathcal{A}^{L,R}_{q}(s,t)\big|_{\rm SM} +
    \dg \frac{\mathcal{R}_{\gamma}}{s}
    +
    \dz \frac{\mathcal{R}_Z^{L,R}}{s-m_Z^2}\,,
    \qquad 
    \mathcal{A}^{L,R}_{d}(s,t)  =
    \mathcal{A}^{L,R}_{d}(s,t) \big|_{\rm SM}\,.
    \label{eq:kappa}
\end{align}
We stress that the $\kappa_{V\gamma}$ factors multiply only the physical pole contributions. Rescaling the entire off-shell effective vertices would lead to a gauge-dependent result, and would incorrectly attribute part of the non-pole amplitude to the pseudo-observables. 

\begin{figure}[t]
 \centering
\hspace{1mm}\includegraphics[width=0.9\textwidth]{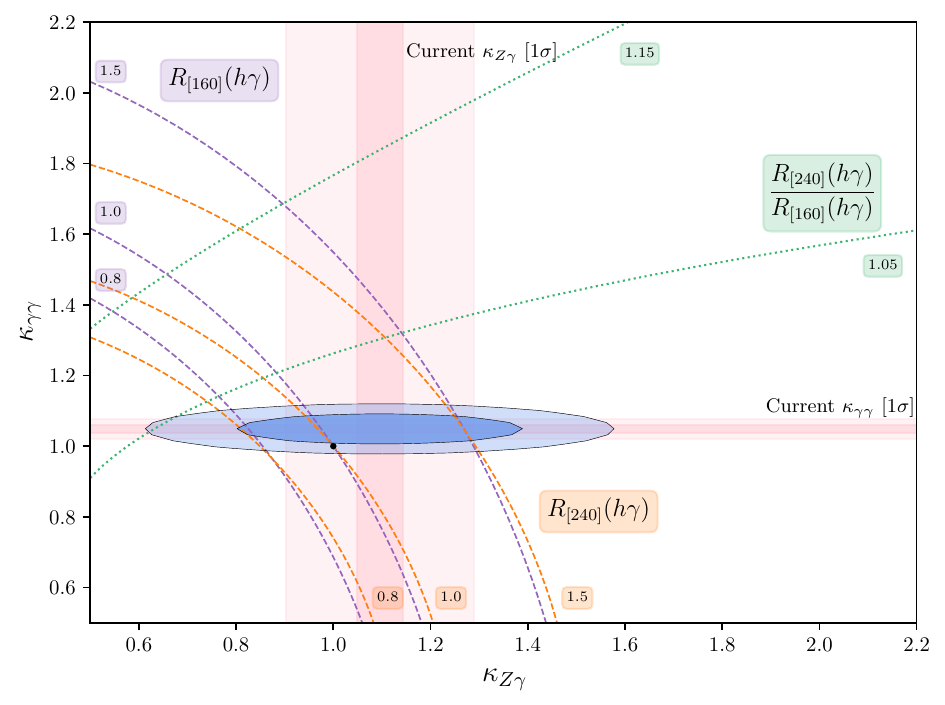}
    \caption{Integrated cross section in the $\kappa_{Z\gamma}$--$\kappa_{\gamma\gamma}$ plane. The shaded regions show the constraints from current determinations of the two radiative Higgs decay rates together with the future FCC-ee sensitivity. The contours indicate constant values of $R_{[160]}(h\gamma)$ (purple dashed), $R_{[240]}(h\gamma)$ (orange dashed), and the corresponding double ratio (green dotted).}
    \label{fig:sigmaexpectA}
\end{figure}

Within the parametrization in Eq.~\eqref{eq:kappa}, angular integration yields an exact quadratic polynomial in the two real modifiers. Defining  
\begin{equation}
R_{[E]}(h\gamma)=\left. \frac{ \sigma(e^+e^-\to h\gamma)[\dg,\dz] }{ \sigma^{\rm SM}(e^+e^-\to h\gamma)} \right|_{\sqrt{s}=E~{\rm GeV}}\,,
\label{eq:R-ratio}
\end{equation}
 we find
\begin{align}
R_{[160]}(h\gamma)={}&1+0.736\dg+1.495\dz+0.317\dg^2+0.109\dg\dz+0.827\dz^2\,,\label{eq:r160}\\
R_{[240]}(h\gamma)={}&1+0.909\dg+1.454\dz+0.538\dg^2+0.146\dg\dz+0.873\dz^2\,.\label{eq:r240}
\end{align}
The different coefficients in Eqs.~\eqref{eq:r160} and \eqref{eq:r240} reflect the energy-dependent interference among the photon-pole, $Z$-pole, and non-pole amplitudes. 
Since Eq.~\eqref{eq:kappa} defines an exact (gauge-invariant) rescaling of the two physical pole residues, we are allowed to retain the terms quadratic in $\dg$ and $\dz$. This does not amount to including arbitrary effects of higher-dimensional operators. As we shall discuss, keeping the full square is essential when discussing the sign-flipped branches, for which a linearized expansion is manifestly inadequate.

\subsection{Numerical analysis}

Figure~\ref{fig:sigmaexpectA} shows the predictions for the integrated cross section in the $\kappa_{Z\gamma}$--$\kappa_{\gamma\gamma}$ plane and compares them with the present determinations of the two pseudo-observables from radiative Higgs decays. Combining the most recent ATLAS and CMS measurements of the $h\to Z\gamma$ signal strength~\cite{ATLAS2026,CMS2026}, we obtain $\kappa_{Z\gamma}^{\rm exp}=1.095\pm0.194$. For $h\to\gamma\gamma$ we use the PDG average, $\kappa_{\gamma\gamma}^{\rm exp}=1.049\pm0.029$. On the SM-connected branch, the present decay constraints still allow an enhancement of the $e^+e^-\to h\gamma$ cross section of order $50\%$.
\begin{table}[t]
\begin{minipage}{0.45\textwidth}
\centering
\begin{tabular}{c||c|c}
$(\kg,\kz)$ & $R_{[160]}(h\gamma)$ & $R_{[240]}(h\gamma)$ \\ \hline\hline
$(+1,+1)$ & 1 & 1 \\
$(-1,+1)$ & 0.797 & 1.333 \\
$(+1,-1)$ & 1.316 & 1.583 \\
$(-1,-1)$ & 1.550 & 2.450 \\
\end{tabular}
\caption{Predicted $e^+e^-\to h\gamma$ rates relative to the SM for pseudo-observables with SM magnitudes and different sign assignments.}
\label{tab:signs}
\end{minipage}
\hspace{0.5 cm}
\begin{minipage}{0.45\textwidth}
\centering
    \begin{tabular}{cccc}
    \hline
        Coupling & Current & HL-LHC & FCC-ee   \\ \hline
        $\kappa_{\gamma \gamma} (\%)$ & 2.8 & 1.6 & 1.1\\
        $\kappa_{Z \gamma}(\%)$ & 18 & 10 & 4.3  \\ \hline
    \end{tabular}
    \caption{Current and projected relative uncertainties on the two radiative Higgs pseudo-observables. The FCC-ee projections include HL-LHC information.}
    \label{tab:kappa_projections}
\end{minipage}
\end{table}

\begin{figure}[t]
\centering\includegraphics[width=0.8\textwidth]{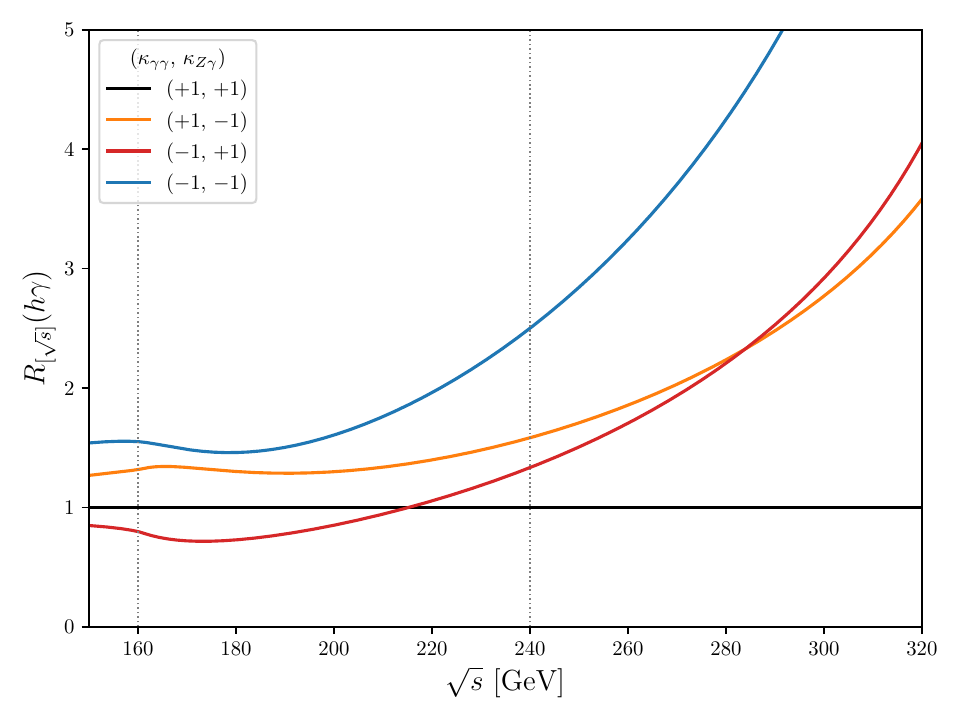}
 \caption{Predictions for $R_{[\sqrt{s}]}(h\gamma)$ as a function of the centre-of-mass energy for $|\kg|=|\kz|=1$ and the four possible sign assignments.}
 \label{fig:signsb}
\end{figure}

Table~\ref{tab:kappa_projections} summarizes the projected precision on the two pseudo-observables from Higgs decays at the end of the HL-LHC program and at FCC-ee~\cite{FCC:2025lpp}. If their central values converge to the SM point, the FCC-ee projections, $\delta\kappa_{\gamma\gamma}=0.011$ and $\delta\kappa_{Z\gamma}=0.043$, restrict the pole-induced variation of $\sigma(e^+e^-\to h\gamma)$ to approximately the $10\%$ level. This is smaller than the anticipated experimental uncertainty of about $30\%$ on the cross section at $\sqrt{s}=240$~GeV. Close to the SM point, the direct $h\gamma$ measurement is therefore not expected to substantially improve the individual precision on $\kappa_{\gamma\gamma}$ and $\kappa_{Z\gamma}$. Its principal value is instead to serve as a consistency test of the pseudo-observable description.

This role becomes particularly clear away from the SM-connected branch (i.e.~for non-positive signs of the $\kappa_{V\gamma}$).
Radiative decay rates determine only $|\kg|$ and $|\kz|$; consequently, even arbitrarily precise rate measurements do not resolve their signs. Table~\ref{tab:signs} gives $R_{[160]}(h\gamma)$ and $R_{[240]}(h\gamma)$ for the four sign assignments obtained by setting $|\kg|=|\kz|=1$. Their distinct predictions show that combining decay information with $e^+e^-\to h\gamma$ measurements can resolve the discrete ambiguity. The use of more than one centre-of-mass energy it is essential to this purpose, allowing to tests the characteristic energy dependence of each sign branch.

More generally, denoting by $\kappa^{\rm obs}_{V\gamma}$ the  absolute values of the $\kappa_{V\gamma}$ determined by Higgs decays,  a useful consistency test is obtained evaluating the four residuals
\begin{equation}
 \Delta_{\pm,\pm}(s) =  \frac{\sigma^{\rm obs} (s)}{\sigma^{\SM} (s)}-
 R_{[\sqrt{s}]}(h\gamma) \big[ \pm\kg^{\rm obs}\,,\, \pm\, \kz^{\rm obs}\, \big]\,.
 \label{eq:null}
\end{equation}
If NP effects are well described by the two Higgs pseudo-observables, one of the four residuals 
must vanish, at every energy. Figure~\ref{fig:signsb} illustrates the four possible values of 
$R_{[\sqrt{s}]}(h\gamma) \big[ \pm\kg^{\rm obs}\,,\, \pm\, \kz^{\rm obs}\, \big]$
for $\kappa^{\rm obs}_{V\gamma}=1$: over the energy range of interest, the fractional separation among the different sign assignments increases with $\sqrt{s}$. A measurement at several energies therefore does more than select a sign branch; it overconstrains the pole-residue hypothesis. If no branch provides a consistent description of the data, the discrepancy would signal non-negligible contact interactions, hence a breakdown of the hypothesis of heavy-NP affecting mainly the pole terms.  

\section{NP effects beyond $\kappa_{\gamma\gamma}$ and 
$\kappa_{Z\gamma}$}
\label{sec:beyond-kappa}
As a prototype of NP effects not captured by modifications of $\kappa_{\gamma\gamma}$ and $\kappa_{Z\gamma}$, we consider a simplified model containing a massive neutral vector field $Z'$. We assume that the $Z'$ couples to electron currents and, through a higher-dimensional interaction, to the Higgs field and the electromagnetic field strength. The relevant $Z'$ interactions terms 
can be parameterized as
\begin{align}
    \mathcal{L}_{Z'} =  \ggp_L (\bar{e}_L \gamma^\mu e_L) Z'_{\mu} + \ggp_R \bar{e}_R \gamma^\mu e_R Z'_{\mu} + \frac{ e \ggp_{h\gamma}} {M_{Z'}^2} Z'_{\mu\nu}F^{\mu\nu} (H^\dagger H)\,.
    \label{eq:LZp}
\end{align}
This simplified model is specified by the four parameters $\ggp_L$, $\ggp_R$, $\ggp_{h\gamma}$, and $M_{Z'}$. By construction, it should be regarded as an illustrative low-energy parametrization rather than a complete UV model.\footnote{$\mathcal{L}_{Z'}$ is written in the electroweak broken phase for the gauge fields, while keeping complete dependence on the Higgs field ($H$). After electroweak symmetry breaking, the last terms induces a $Z^\prime$--$\gamma$ kinetic mixing, whose tree-level effect can be absorbed via a redefinition of the couplings $\ggp_{L,R}$.}
After electroweak symmetry breaking, $Z'$ exchange generates the helicity-conserving structure in Eq.~\eqref{eq:contact-structure}. If $M_{Z'}^2\gg s$, its propagator can be expanded and the leading term is precisely that induced by a dimension-eight SMEFT operator. If instead $M_{Z'}$ lies not far above the collision energy, the full propagator produces a characteristic non-local energy dependence that cannot be absorbed into the two pole pseudo-observables.

We focus on the kinematical regime
\begin{equation}
m_Z^2 \ll s < M_{Z'}^2\,,
 \label{eq:Zp-regime}
\end{equation}
where the $Z'$ cannot be produced on shell, while the collision energy is sufficiently high to enhance its virtual effect. The $Z'$ contribution to the $q$-type scalar amplitudes is
\begin{equation}
\Delta\mathcal A_q^{L,R}(s)\big|_{Z'}=
\ggp_{L,R}
 \frac{2ev\ggp_{h\gamma}}{M_{Z'}^2\left(s-M_{Z'}^2\right)}
 \label{eq:Zp-amplitude}
\end{equation} 
that, as expected,  reduces to a constant term in the limit $s\ll M_{Z'}^2$. 

The limit $s\gg m_Z^2$ leads to an important simplification on the SM pole amplitude, which 
becomes approximately left handed. Indeed, the residues reported in the appendix satisfy $\mathcal{R}^R_Z\approx-\mathcal{R}_\gamma$, implying an approximate cancellation of the vector-like  $\gamma$- and $Z$-pole terms when $m_Z^2$ is neglected in the propagator:
\begin{align}
    \ \mathcal{A}^{L}_{q}(s)\big|^{\rm SM}_{\rm pole} & \approx \frac{1}{s} 
    ( \mathcal{R}_Z^{L} +   \mathcal{R}_\gamma)  \approx 6.3  \times \frac{e g^3 m_Z}{16 \pi^2 c_W} \frac{1}{s\,m_H^2}\,,\label{eq:SM-high-energy}    \\
        \ \mathcal{A}^{R}_{q}(s)\big|^{\rm SM}_{\rm pole} & \approx \frac{1}{s} 
    ( \mathcal{R}_Z^{R} +   \mathcal{R}_\gamma)  \approx 0\,.
\end{align}
The $Z'$ contribution proportional to $\ggp_L$ can therefore interfere with the dominant SM amplitude, whereas the effect proportional to $\ggp_R$ plays a subleading role. Combining Eqs.~\eqref{eq:Zp-amplitude} and \eqref{eq:SM-high-energy}, we get
\begin{align}
     \mathcal{A}^{L}_{q}(s)\big|^{{\rm SM}+Z'}_{\rm pole} &\approx   
     \mathcal{A}^{L}_{q}(s)\big|^{{\rm SM} }_{\rm pole} \left[ 1 +  0.62  \times \frac{16\pi^2 c_W^2\,\ggp_L \ggp_{h\gamma }}{g^4} \frac{m_h^2}{ M^2_{Z'}}  \frac{ s}{ s-M^2_{Z'}} \right]\\
     &\approx   
     \mathcal{A}^{L}_{q}(s)\big|^{{\rm SM} }_{\rm pole} \left[ 1 +  0.35  \times \left(
    \frac{\delta_Z}{10^{-3}}\right)
    \frac{\ggp_{h\gamma}}{ \ggp_L }
     \frac{ s}{ s-M^2_{Z'}}  \right]\,,
     \label{eq:AZp}
\end{align}
where we have defined 
\begin{equation}
    \delta_Z =  \left( \frac{ \ggp_L m_Z }{ g M_{Z'}} \right)^2\,. 
    \label{eq:deltaZ}
\end{equation} 
The parameter $\delta_Z$ is the ratio of  $(\ggp_L)^2/M_{Z'}^2$, namely the effective scale of the $Z'$-induced four-electron contact interaction, and the Fermi scale. It is therefore a convenient proxy for the sensitivity of precision $e^+e^-\to e^+e^-$ measurements to the $Z'$ exchange. Current electroweak measurements imply $\delta_Z\lesssim10^{-3}$, but the $Z$-pole run at the FCC-ee should push the sensitivity on $\delta_Z$ well below the $10^{-4}$ level.

Pulling out the explicit dependence from 
$\delta_Z$, 
Eq.~\eqref{eq:AZp} illustrates the non-trivial conditions necessary to have a sizable $Z'$ impact on $e^+e^- \to h \gamma$. This 
 requires both a strong hierarchy $\ggp_{h\gamma}\gg\ggp_L$ and a $Z'$ mass close enough to the accessible kinematic range to benefit from the propagator enhancement. The first condition is highly non-generic in UV completions. Moreover, even if both conditions are met, the same state would generally be more readily probed through precision tests at the $Z$ pole. 

This simplified example illustrates a general conclusion: once the $h\gamma\gamma$ and $hZ\gamma$ pole residues are constrained by Higgs decays, producing a large additional effect in $e^+e^-\to h\gamma$ is difficult if all new states are beyond the kinematical reach. An observable departure from the pseudo-observable prediction would point to both unusually hierarchical couplings and new states sufficiently close to the collision energy that their non-local propagator structure becomes evident by the
energy dependence of the cross-section. 

\section{Conclusions}

The process $e^+e^-\to h\gamma$ provides an interesting  probe of the effective couplings of the Higgs boson to neutral gauge bosons. We have revisited the complete SM one-loop calculation and organized the amplitude into separately gauge-invariant 
$\gamma$- and $Z$-pole terms, and non-pole contributions. This decomposition provides a very useful basis to analyse NP effects. As we have shown, for generic heavy new states, power counting implies that the leading deviations from the SM are necessarily encoded in modifications of the $h\gamma\gamma$ and $hZ\gamma$ pole residues. These two pseudo-observables, which are  also accessible in Higgs decays, provide a robust, model-independent framework for interpreting future $e^+e^-\to h\gamma$  measurements.

Interestingly enough, this process contains information that cannot be obtained from the corresponding Higgs decay rates: measurements of $e^+e^-\to h\gamma$ at different centre-of-mass energies can resolve the unavoidable sign  ambiguity on the $h\gamma\gamma$ and $hZ\gamma$ effective couplings resulting from Higgs decays. For small deviations from  the SM values, the sensitivity to the two pseudo-observables expected from $e^+e^-\to h\gamma$ 
does not improve the one derived  from  Higgs decays. However, measuring  $\sigma(e^+e^-\to h\gamma)$ at $\sqrt{s}=160$~GeV and $\sqrt{s}=240$~GeV 
at FCC-ee would
allow us to select the correct sign branch and to over-constrain the  hypothesis of NP encoded only via the 
Higgs pseudo-observables. 

Finally, we have analysed under which  conditions the description of this process only via the two Higgs pseudo-observables could fail. By means of a simplified $Z'$ model, we have shown that detectable NP effects other than those encoded in the $h\gamma\gamma$ and $hZ\gamma$ effective couplings are very difficult to generate assuming heavy NP.
They require new states close to the kinematical threshold with rather unnatural couplings. A pattern of 
$\sigma(e^+e^-\to h\gamma)$ at different $\sqrt{s}$ values inconsistent with all sign assignments for the two Higgs pseudo-observables would point to new degrees of freedom sufficiently light and strongly coupled that the description of NP effects in terms of local operators ceases to be adequate.

\subsection*{Added Note}
While this paper was being completed, a detailed estimate of the FCC-ee sensitivity in measuring $\sigma(e^+e^-\to h\gamma)$, at different energies, has been presented~\cite{Herrmann:2026exd}. The
results of Ref.~\cite{Herrmann:2026exd}
provide a solid basis to confirm our qualitative conclusion that FCC-ee will resolve the sign ambiguities in the $h\gamma\gamma$ and $hZ\gamma$ effective couplings resulting from Higgs decays. On the other hand, we stress the importance of analysing future $e^+e^-\to h\gamma$ data  via the Higgs pseudo-observables framework proposed in this work, which provides a complementary and more general approach to the one based on specific sets of Wilson coefficients discussed  in~Ref.~\cite{Herrmann:2026exd}. 

\subsection*{Acknowledgments}
We are grateful to Admir Greljo  and Michele Selvaggi for useful discussions. We also thank Marumi Kado for asking questions that prompted us to start this project. The work is supported by the Swiss National Science Foundation, project No.~2000-1-240011, and by the Swiss High Energy Physics initiative for the FCC (CHEF).

\appendix

\section{Analytic Expressions of Form Factors}

Below, we list the analytic expressions for the subsets of diagrams for the process \(e^+ (p_+) e^- (p_-) \rightarrow \gamma (k) h (p_h)\) described in Eq.~\eqref{eq:Mtot}. We define $q^2 = (p_+ + p_-)^2=s$ and use
\begin{align}
    C_0^\chi (s) &= C_0(0, s, m_H^2, m_\chi^2, m_\chi^2, m_\chi^2) \\
    C_{12}^\chi (s) &= C_{12} (0, m_H^2, s, m_\chi^2, m_\chi^2, m_\chi^2)\\
    D_{n}^W (s,x) &= D_n(0, m_H^2, 0,0,x,s,0, m_W^2, m_W^2, m_W^2) \\
    D_{n}^Z (x_1,x_2) &= D_n( 0,0,m_H^2, 0, x_1, x_2, 0,0,m_Z^2, m_Z^2 )
\end{align}
with the three- and four-point functions 
($C_{n}$ and $D_{n}$) defined as in \cite{Denner:1991kt}.

\subsection{s-channel Photon}
\begin{align}
    \mM_\gamma &=  \bar{v}(p_+)\left[ \,e\, Q_e\; \gamma^\alpha G^\gamma_{\alpha\mu}(s)\;i \Gamma_{h \gamma \gamma}^{\mu\nu}(s) \right]u(p_-)\epsilon_\nu^*(k) \\
    \Gamma_{h \gamma \gamma}^{\mu\nu} (s) &= \frac{ \,e^2 g}{16 \pi^2 }m_W
     \left[ \left(\mathcal{F}_\gamma^W(s) -\sum_f 4 Q_f^2  N_c \frac{m_f^2}{m_W^2} \mathcal{F}_\gamma^f(s)\right)  \mathcal{T}_{\mu\nu}(q,k)  + \mathcal{G}_\gamma^W (s) g^{\mu \nu}  \right]  \\
    \mathcal{F}^f_\gamma(s)&=(C^f_0(s) - 4 C^f_{12}(s))\,\\
    \mathcal{F}_\gamma^W(s) &=\left(16 C_0^W(s) -4 \left(\frac{m_h^2}{m_W^2}+6 \right) C_{12}^W(s) \right)  \\     
       \mathcal{G}^W_\gamma(s)  &= - 3 C_0^W(s)\; s  \\
    \mathcal{R}_\gamma &= \frac{e^3 Q_e g}{16 \pi^2} m_W  \left(\mathcal{F}_\gamma^W(0) -\sum_f 4 Q_f^2  N_c \frac{m_f^2}{m_W^2} \mathcal{F}_\gamma^f(0)\right) \nonumber \\
    &= -15.76 \times \frac{e^3 Q_e g}{16 \pi^2} \frac{m_W}{m_H^2}
\end{align}
\subsection{s-channel Z}

\begin{align}
    \mM_Z &=\bar{v}(p_+)\left[ \frac{g}{c_W} \gamma^\alpha\left(P_L \,T^3_e -Q_e s_W^2\right)G^Z_{\alpha\mu}(s) \;i\Gamma_{ h Z\gamma}^{\mu\nu}(s)\right]u(p_-)\epsilon_\nu^*(k) \\
     \Gamma_{h Z \gamma}^{\mu\nu} (s) &= \frac{\,e\, g^2}{16 \pi^2 }m_Z \left[ \left(\mathcal{F}_Z^W(s) + \sum_f \frac{m_f^2}{m_W^2}  2\,N_c Q_f (2 \,Q_f s_W^2 - T^3_f) \mathcal{F}^f_Z(s) \right)\mathcal{T}_{\mu\nu}(q,k) +  \mathcal{G}_Z^W (s) g^{\mu \nu} \right] \\ 
     \mathcal{F}^f_Z(s)&=(C^f_0(s) - 4 C^f_{12}(s)) \\
     \mathcal{F}^W_Z(s)&=  4 \left(4 c_W^2-1\right)C_0^W(s)+2 \left(2 \left(1-2 c_W^2 (D-1)\right)+\frac{\left(1-2 c_W^2\right) m_h^2}{m_W^2}\right) C_{12}^W(s)  \\
     \mathcal{G}_Z^W(s) 
   &=  -3 c_W^2 C_0^W(s) (s-m_Z^2) \\
   \mathcal{R}^L_Z &= \frac{e g^3 m_Z}{16 \pi^2 c_W} (T_e^3 -Q_e s_W^2) \left(\mathcal{F}_Z^W(m_Z^2) + \sum_f \frac{m_f^2}{m_W^2}  2\,N_c Q_f (2 \,Q_f s_W^2 - T^3_f) \mathcal{F}^f_Z(m_Z^2) \right) 
   \nonumber\\
    &= -13.06 \times \frac{e g^3 m_Z}{16 \pi^2 c_W} (T_e^3 -Q_e s_W^2) \frac{1}{m_H^2}  \\
   \mathcal{R}^R_Z &=- \frac{e g^3 m_Z Q_e s_W^2}{16 \pi^2 c_W} \left(\mathcal{F}_Z^W(m_Z^2) + \sum_f \frac{m_f^2}{m_W^2}  2\,N_c Q_f (2 \,Q_f s_W^2 - T^3_f) \mathcal{F}^f_Z(m_Z^2) \right) \nonumber\\
    &=+13.06\times \frac{e g^3 m_Z Q_e s_W^2}{16 \pi^2 c_W} \frac{1}{m_H^2} 
\end{align}

\subsection{W Boxes}

\begin{align}
    \,\mathcal{M}_{W \text{-box}}&= \bar{v}(p_+) \gamma^\mu P_L\, \mathcal{B}^{ W }_{\mu\nu}   (s,t,u)\mathit{u}(p_-) \epsilon^\nu_\gamma \\
   \mathcal{B}^{W}_{\mu\nu}(s,t,u) &=\,\frac{e g^3 m_W}{16\pi^2} \left[\mathcal{T}_{\mu\nu}(p_-,k) F_{W\text{-box}}(s,t,u)+\mathcal{T}_{\mu\nu}(p_+,k) F_{W\text{-box}}(s,u,t)+    \mathcal{G}^W_{B}(s) g_{\mu\nu}\right]\\
    \mathcal{G}^W_{B}(s)  &=  T_e^3\,e\, g^3 \, m_W \,C_0^W(s)  \\
       F_{W\text{-box}}(s,x_1,x_2) &= (2-D)\,( D_{23}^{W}(s,x_1)+ D_{22}^{W}(s,x_1)- D_{12}^{W}(s,x_2)) \nonumber \\
    &\qquad-(D\,-4) D_2^{W}(s,x_1)+2 ( D_3^{W}(s,x_1)+ D_1^{W}(s,x_2)) 
\end{align}

\subsection{Z Boxes}

\begin{align}
    \mathcal{M}_{Z\text{-box}} &= \bar{v}(p_+) \gamma^\mu \frac{1}{c_W^3}  \left(  \frac{1}{2}(1-2s_W^2)^2P_L+2 s_W^4P_R   \right)\,\mathcal{B}^{Z}_{\mu\nu}(s,t,u)u(p_-) \epsilon^\nu_\gamma \\
  \mathcal{B}^{Z}_{\mu\nu}(s,t,u)&=\frac{e g^3 m_Z}{16\pi^2}\left[\mathcal{T}_{\mu\nu}(p_-,k) F_{Z\text{-box}}(t,u)+\mathcal{T}_{\mu\nu}(p_+,k) F_{Z\text{-box}}(u,t) \right] \\ 
    F_{Z\text{-box}}(x_1,x_2) &=(D\,-2)( D_{13}^Z(x_2,x_1) + D_{23}^Z(x_2,x_1))+2 D_3^Z(x_2,x_1)
\end{align}

\bibliographystyle{JHEP}
\bibliography{bibliography.bib}

\end{document}